# On the Observability and Redundancy of Intelligent Transportation Networks

**Mohammadreza Doostmohammadian**[1,*]

1 Mechatronics Group, Faculty of Mechanical Engineering, Semnan University, Semnan, Iran; doost@semnan.ac.ir

**Abstract:** The safety and reliability of intelligent-transportation-systems (ITS) can be greatly enhanced through adding redundancy in the information-sharing network of the vehicles. In this paper, we first model the mixed traffic of human-driven and autonomous vehicles as a distributed system observability problem over a network of communicating vehicles. We clearly show that a strongly-connected network with minimum $n$ links (with $n$ as network size) is sufficient for the observability of mixed traffic network. Then, we present graph-theoretic results on adding redundancy to the changing network of vehicles to make it resilient to the failure of certain number of vehicles/sensors or their data-sharing links. Finally, we employ distributed observer design to validate our results over a simple mixed traffic example.

## 1. Introduction

Redundant design of the dynamical systems and automatic control systems [1,2] has been the topic of signal processing and control literature. These systems, which are related to many applications—including industrial automation, robotics, sensor networks, and smart transportation—must not only perform their intended functions properly but also withstand and recover from failures. Redundancy, as a fundamental design principle, plays a key role in enhancing the safety and robustness of these systems especially in large-scale applications. By incorporating multiple components or pathways that can perform similar functions, redundancy ensures that a system can continue operating properly even when a failure occurs in a part of its architecture. This is, for example, by adding equivalent measurements [3], redundant sensors [4,5], and extra linkings over the connecting network [6,7]. The redundant design is also considered for reducing criticality of road links [8], planning of road network junctions [9,10], identifying the critical stations of metro networks [11,12], and improving the resilience of transportation systems [13,14]. There are practical challenges of simulation and validation for connected automated vehicles. The work [15] discusses some existing challenges on integrated simulation environments for connected and automated vehicles.

It should be noted that adding redundancy results in a trade-off between robustness and complexity. While redundancy can indeed increase robustness in distributed systems, in complex cyber-physical infrastructures such as transportation networks it may also introduce additional system complexity and interactions between subsystems that do not always complement each other. Redundant sensing or communication channels, for instance, may produce inconsistent measurements, synchronization issues, or operational conflicts if integration between systems is not carefully designed. In other words, redundancy improves fault tolerance and observability coverage, yet in cyber-physical transportation networks with mixed traffic and heterogeneous sensing/communication stacks, it can create coordination challenges, measurement inconsistencies, and integration risks that must be managed through careful system design.

Observability, the ability to infer the internal state of a system from its external outputs, complements redundancy by enhancing the system's capacity to detect and respond to anomalies, faults, or attacks [16]. Observability plays a key role in the distributed estimation of large-scale dynamical systems [17] or time-varying networks [18]. Together, redundancy and observability form a robust framework for safer and more reliable operations in distributed control systems, including ITS. This combination not only facilitates fault detection and diagnosis [16] but also enables measures to mitigate potential disruptions before they escalate into critical failures [19]. The principles of redundancy and observability can be particularly transformative when applied to the inference of ITS. In these complex systems, where real-time decision-making and coordinated actions are vital, the integration of redundant channels and observability perspective [20] can considerably improve resilience against failures caused by traffic incidents, infrastructure breakdowns, or malicious attacks. As *intelligent transportation* demands increase, the need for systems that can presume working (with safety) in the face of uncertainties and faults is of great interest. In this direction, this paper explores the role of the observability redundancy within ITS. In particular, we will examine how these concepts can be extended to transportation networks modelled as dynamical systems. Recall that dynamical network modelling accounts for both node (vehicle) dynamics and structure of the ITS network (as a graph) [21]. Using this model, the current work provides a technique toward the development of more resilient and robust transportation setups, for example, in vehicle platooning [22]. In the ITS setup, this also finds application for safe merging scenarios of autonomous vehicles (AVs) and human-driven vehicles (HDVs) in the mixed traffic [23-25].

In this paper, we clearly formulate the mixing traffic scenario as a distributed observability problem in system-theoretic perspective where both linear system models and network-theoretic notions are involved. We develop the notion of distributed observability of the network of mixed traffic (both HDVs and AVs) and mathematically formulate this via algebraic terms. As our main contribution, introducing the notion of system graph observability, we derive sufficient conditions for distributed observability of the group of HDVs by the network of AVs. Then, we provide a distributed observer scenario that only needs the strong-connectivity of the AV network for tracking HDV states. This significantly reduces the connectivity requirement and network traffic as compared to many existing literature requiring more than strong-connectivity [26-31]. This further allows redundant design for resilient tracking. We discuss relevant graph-theoretic notions and propose an algorithm for the redundant design of the information-sharing network of AVs such that the transportation network is resilient to certain number of link (or node) failures (time-varying network setup). To validate the results, we perform simulations on a sample mixed HDV-AV traffic network under node and link failure.

The rest of the paper is organized as follows. In Section 2 we formulate the problem of intelligent transportation in the observability perspective. Section 3 presents the results on distributed observability over transportation networks. Section 4 provides the algorithms on the redundant networked observer design. Section 5 gives an illustrative example. Finally, Section 6 concludes the paper.

## 2. Statement of the Problem

In this section, we first derive the conditions for the observability of the HDVs by the connected network of AVs. This is necessary to estimate the state of the HDVs for merging scenarios and stable mixed traffic of HDVs and AVs. Consider a group of AVs to move autonomously within the HDV-AV traffic as in Fig. 1. Among the group of AVs some take *direct sensor measurement* of the state of the HDVs and the rest obtain this information via the *information-sharing network* of AVs, denoted by $\mathcal{G}_W$. Let the state dynamics of HDV $i$ be in the following state-space model:

$$x_{i,k+1} = A_i x_{i,k} + B_i u_{i,k}, \tag{1}$$

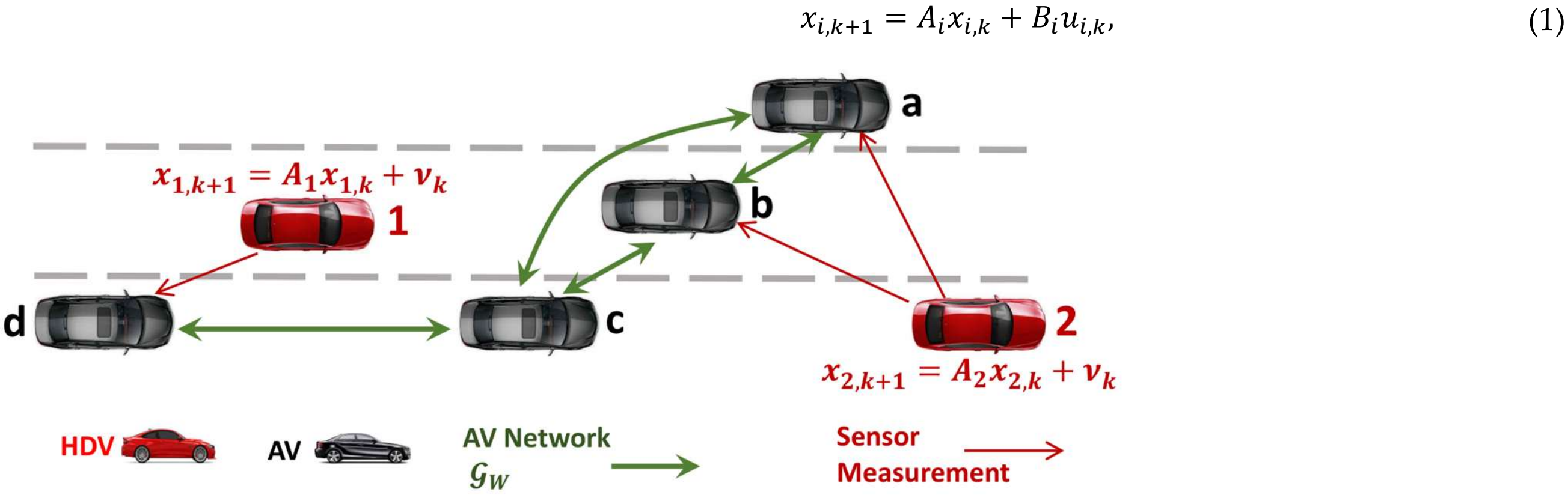

**Figure 1.** A mixed traffic scenario with both AVs and HDVs: for stable traffic movement the states of the HDVs need to be measured and tracked by the AVs. Vehicle '**c**' tracks the state of HDV number 2 via information-sharing with both vehicles '**a**' and '**b**'. In case the link '**bc**' fails, the *redundant* link '**ac**' delivers the state information to vehicle '**c**' for stable tracking. On the other hand, if the link '**cd**' fails, there is no other way to deliver state information of the dynamics of HDV number 1 to the rest of the AVs.

with state $x_{i,k+1} \in R^m$ and input $u_{i,k} \in R^{l_i}$ at time-step $k$. Matrix $A_i \in R^{m\times m}$ and $B_i \in R^{m\times l_i}$ as system and input matrix. One can concatenate the state of all $N$ HDVs and the associated sensor measurements in a compact form as

$$x_{k+1} = Ax_k + Bu_k, \tag{2}$$

with $x_k = \left[x_{1,k}, \dots, x_{N,k}\right]^\top \in R^{Nm}$ as the states of all HDVs at time $k$. The block diagonal matrix $A = \text{blockdiag}[A_i] \in R^{Nm\times Nm}$ and $B = \text{blockdiag}[B_i] \in R^{Nm\times Nl_i}$ are the overall state-space system and input matrices. We consider *discrete-time* dynamical systems, and it is known that such discretization models of continuous-time dynamical systems result in (generically) full-rank system matrix $A$ with nonzero diagonal entries (referred to as self-damped), see details in [32]. Furthermore, most existing tracking models for example nearly-constant-velocity (NCV), nearly-constant-acceleration (NCA), and Singer dynamics [33] are full-rank. The HDV model is associated with sensor measurements by AVs in the form:

$$y_k = Cx_k + \mu_k, \tag{3}$$

where $y_k = \left[y_{1,k}, \dots, y_{n,k}\right]^\top \in R^l$ and $\mu_k$ as the measurement vector and output noise at time $k$. The output matrix $C \in R^{l\times Nm}$ is the output matrix. It should be mentioned that *not all the AVs take state measurement of the HDVs*. Only some of the AVs take direct sensor measurements of the states of the nearby HDVs and, then, the information is shared over the network $\mathcal{G}_W$. This is also shown in Fig. 1. For stable mixing traffic, the unknown state of the HDVs must be estimated by the rest of the AVs and this is the place where observability plays the key role. In other words, for the rest of AVs with no direct measurement, the state of HDVs must be *observable*. This is achieved via the information-sharing network of AVs, denoted by $\mathcal{G}_W$, and is referred to as *distributed observability* or *networked observability*. This implies that *no local observability requirement* at any vehicle is assumed, instead distributed observability is obtained via the information-sharing network of the AVs and applying proper networked estimation protocol. The problem in this paper is to first formulate distributed observability; then, derive the sufficient conditions on $\mathcal{G}_W$ to satisfy distributed observability, and, finally, propose *redundant* design of the $\mathcal{G}_W$ such that the distributed observability holds in spite of *failure* in some of the AVs or their communication links over $\mathcal{G}_W$. The distributed observability conditions and redundant design are based on some *graph-theoretic* concepts described in the next section. Fig. 2 visually summarizes the problem statement, objectives and methodology.

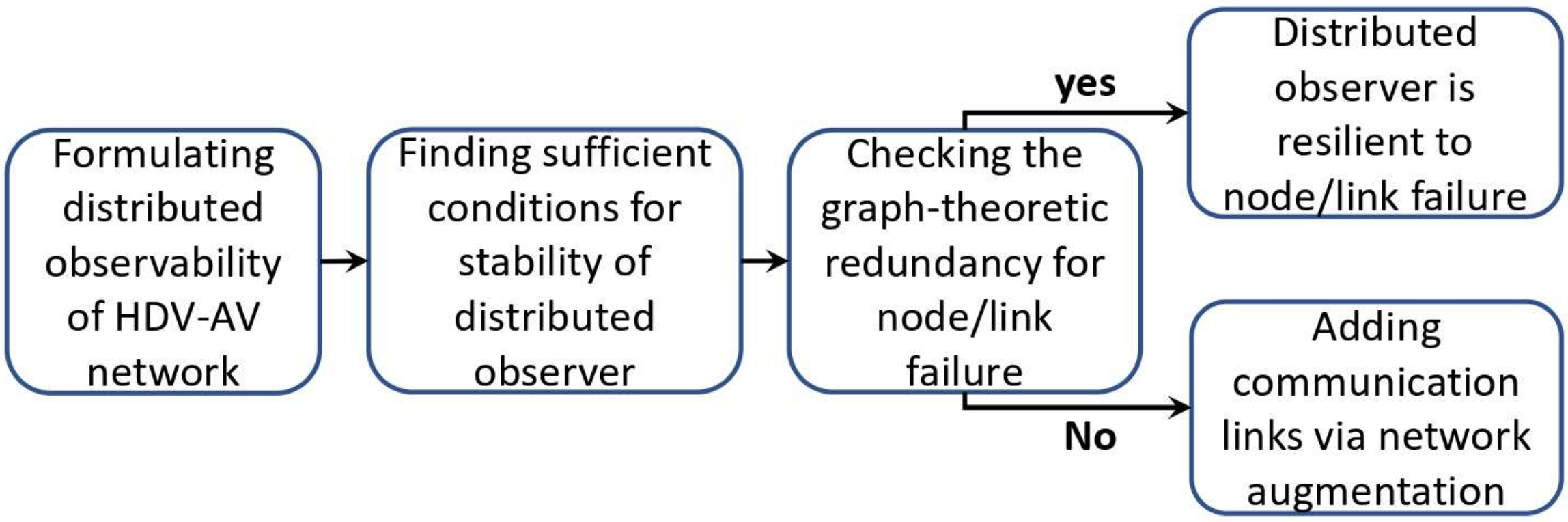

**Figure 2.** The flowchart summarises the objectives and methodology in this paper.

## 3. Distributed Tracking via Observability on Graphs

In networked control systems and sensor networks, it is typical to analyze the linear system-observation model by (2) and (3) in graph-theoretic perspective. In the system graph representation $\mathcal{G}_A$, the states are modelled as nodes and every nonzero entry of $A$, say $a_{ij}$, represents a link connecting state node $j$ to $i$. Similarly, the entry $C_{ij}$ of the observation matrix $C$ represents a link from the state node $j$ to output node $i$. In this direction, the observability can be analyzed via graph-theoretic terms as given in the next lemma.

***Lemma 1.*** *[34] Given a full-rank system matrix $A$, the pair $(A, C)$ is observable if for every state node $i$ in $\mathcal{G}_A$ there exists a directed path starting from node $i$ and ending to an output node $j$.*

The condition in Lemma 1 is known as *output-connectivity*. In the distributed setup, we need to extend the output-connectivity to the *global* system dynamics. In the global model the information of every AV node $i$ is shared with its directed neighbours $j \in N_i$ over $\mathcal{G}_W$. This global state dynamics is modelled by the Kronecker product of the system matrix $A$ and $W$ matrix as the stochastic adjacency matrix of the network $\mathcal{G}_W$ as [35],

$$\underline{x}_{k+1} = (W \otimes A)\underline{x}_k + 1_n \otimes u_k, \tag{4}$$

with $\otimes$ denoting the Kronecker product. Note that, in this setup, the entry $w_{ij}$ of the adjacency matrix $W$ represents the weight assigned to the information received to AV $i$ from AV $j$. This represents a *consensus-type* setup with weight-matrix $W$ being stochastic, i.e., $\sum_{j=1}^{n} w_{ij} = 1$ [36]. This implies an averaging-type dynamics at AVs to filter information received from their neighbours. The diagonal entries $w_{ii}$ are all non-zero, implying that every AV uses its own information (a-priori estimates and/or measurements) to update its observed state.

As an example, consider the single-time-scale distributed estimator/observer. The dynamics include consensus on a-priori estimates and innovation-update based on measurements, and is formulated as follows:

$$\hat{x}_{i,k|k-1} = \sum_{j \in N_i} w_{ij} A \, \hat{x}_{j,k|k-1}, \tag{5}$$

$$\hat{x}_{i,k|k} = \hat{x}_{i,k|k-1} + K_i \sum_{j \in N_i} C_j^T \left( y_{j,k} - C_j \hat{x}_{j,k|k-1} \right), \tag{6}$$

with $K_i$ as the local gain matrix, and $\hat{x}_{i,k|k-1}$ and $\hat{x}_{i,k|k}$ as a-priori and posteriori estimates at AV $i$. The global error dynamics (concatenated at all AVs) for this distributed estimator follows as[1]

$$\underline{e}_{k+1} = \left( W \otimes A - K D_{C(W \otimes A)} \right) \underline{e}_k + \zeta_k, \tag{7}$$

with $\zeta_k$ collecting the noise and input terms and $D_C = \text{blockdiag}\left[ \sum_{j \in N_i} C_j^{\top} C_j \right]$. Then, for Schur stability of the error dynamics (7) we need the pair $(W \otimes A, D_C)$ to be observable, which follows the Kalman stability theorem [28].

***Remark 1.*** *The observability of the pair* $(W \otimes A, D_C)$ *denotes the mathematical formulation for distributed observability.*

***Theorem 1.*** *Let the conditions in Lemma 1 for observability of the system graph* $\mathcal{G}_A$ *holds. If the network* $\mathcal{G}_W$ *connecting the AVs is strongly-connected (SC), then* $(W \otimes A, D_C)$*-observability holds.*

***Proof.*** *First, we prove that matrix* $W \otimes A$ *is full-rank. For an SC network* $\mathcal{G}_W$ *its adjacency matrix* $W$ *is irreducible [38]. Recall that in this irreducible structure, all the diagonal entries are nonzero. This implies that the system is self-damped and, thus, is structurally full-rank [39]. Then following the properties of the Kronecker product, we know that the Kronecker product of two full-rank matrices is also full-rank [40]. Next, following Lemma 1, we check that every state node in the global system* $W \otimes A$ *is output-connected. Given that the pair* $(A, C)$ *is observable, all its state nodes are output-connected from Lemma 1. Note that the system graph* $\mathcal{G}_A$ *can be decomposed into strongly-connected-components (SCCs) and, accordingly, the matrix* $A$ *can be decomposed into irreducible structures each associated with an SCC. Note that, from Lemma 1, the output-connectivity implies that every SCC is connected to an output. Next, following the definition, if the network* $\mathcal{G}_W$ *is SC then there is a directed path from every node to every other node in the graph [40]. This implies that in the composite graph* $\mathcal{G}_W \times \mathcal{G}_A$ *(associated with the global system* $W \otimes A$*) there are giant SCCs made from the Kronecker product of the SC network* $\mathcal{G}_W$ *and SCCs in* $\mathcal{G}_A$*. Note that these giant SCCs contain the main smaller SCCs in* $\mathcal{G}_A$ *which are primarily output-connected. Following the definition of SCC, there is a path from every node in the giant SCC to this output-connected subgraph and, thus, all nodes in every giant SCC in* $\mathcal{G}_W \times \mathcal{G}_A$ *are output-connected. This proves the output-connectivity of the composite graph replicating the structure of the global system* $W \otimes A$*. It should be noted that the observability of the Kronecker product system* $W \otimes A$ *follows the structure of the system graph associated with* $A$ *and network associated with* $\mathcal{G}_W$*. This complex network can be seen as a network of networks or a Kronecker product network associated with the* $W \otimes A$ *matrix. To analyze the observability of this Kronecker product network recall some results from Ref. [41]. Ref. [41] clearly proves the observability condition for the Kronecker product network in the form* $\mathcal{G}_W \times \mathcal{G}_A$ *with the adjacency matrix in the form* $W \otimes A$*. Given that* $A$ *is full-rank and minimal measurements for output-connectivity are given in*

[1] See detailed derivation of the error dynamics in [37]

*Lemma 1, Theorem 4 in [41] gives the (minimal) sufficient conditions for output-connectivity of $\mathcal{G}_W \times \mathcal{G}_A$ as network $\mathcal{G}_W$ being strongly-connected and self-damped. Recall that $\mathcal{G}_W$ is self-damped because every (AV) node uses its own information over the network and thus all the diagonal entries of $W$ are nonzero. Therefore, strong-connectivity of $\mathcal{G}_W$ is sufficient for output-connectivity and $(W \otimes A, D_C)$-observability.*

Recall that $(W \otimes A, D_C)$-observability implies that the distributed observer can locally track the state of the system dynamics given by Eq. (2) with steady-state stability (via the proper design of the feedback gain matrix $K_i$). Then, the following corollary holds from Theorem 1.

***Corollary 1.*** *Assume that there is one sensor measurement from every HDV with dynamics $A_i$ to (at least) one AV, as in Fig. 1. Using distributed observer (5)-(6) and designing the information-sharing network $\mathcal{G}_W$ to be SC, $(W \otimes A, D_C)$-observability holds. This implies that the states of all HDVs are observable to all AVs for tracking purposes. This is better illustrated by the example in Section 5.*

## 4. Redundant Design

Recall that, as proved in Theorem 1, the network requirement for our distributed observer is relaxed to strong-connectivity. This allows for the redundant design to add extra links to the network of AVs to make it more robust to link/node failure. As compared to many distributed observers in the literature, this redundant network design and relaxed network requirement is a privilege towards resilient tracking over time-varying networks, see Table 1 for better comparison. Note that the works [26-31] cannot address redundant design as they require more than strong-connectivity.

**Table 1.** Comparison of the proposed distributed observer in terms of network-connectivity × communication-rate per iteration of system dynamics.

| Ref. | minimum required links × communication rate |
|---|---|
| [26] | $n(n-1) \times 1$ |
| [27,28,29] | $3n \times 1$ |
| [30,31] | $n \times L$ with $L$ more than network diameter |
| This work | $n \times 1$ |

The primary goal of this section is to provide a network topology that can sustain connectivity even when certain links or nodes are compromised or removed. A key concept in this domain is the idea of $\kappa$*-link-connected* and $\kappa$*-node-connected* graphs. A $\kappa$*-link-connected* (k-node-connected) graph/network is a SC graph that remains connected even after the removal of any $\kappa - 1$ links (nodes). This is by increasing the *path multiplicity* among nodes that enhances the resilience of the network and allows for alternative routes of data-sharing to exist, thus mitigating disruption caused by failures. Here, we provide an algorithm for redundant network design[2] We present a graph *augmentation* method (see Algorithm 1 below) which aims to add additional links to a given graph to heuristically achieve $\kappa$*-link-connectivity*. Other similar methods such as the greedy-type algorithms [43] can be employed to identify the minimum number of (or cost-optimal) links to add.

Algorithm 1 takes a (seed) graph $\mathcal{G}_W = (\mathcal{V}, \mathcal{E})$ and an integer $\kappa$. For each node in the graph, it finds additional nodes to connect to ensure the degree of all nodes reaches $\kappa$ with complexity $\mathcal{O}(|\mathcal{V}|)$. This is because for $\kappa$-link connectivity the minimum node degree must be $\kappa$ [44]. Then, connected components are defined via Tarjan or DFS algorithm with complexity $\mathcal{O}(|\mathcal{V}| + |\mathcal{E}|)$. The algorithm checks for disconnected components and adds links between them to reach κ-link-connectivity. This depends on the graph structure and in the worst case could be $\mathcal{O}(|\mathcal{E}|)$. Based on this the overall complexity is $\mathcal{O}(|\mathcal{V}| + |\mathcal{E}|) = \mathcal{O}(|\mathcal{V}|^2)$ and is thus scalable. However, the algorithm outcome may not be the most efficient and optimal solution. Recall that the optimal network augmentation for κ-link connectivity is generally NP-hard [45-47]. For example, for comparison, the algorithm in [45] is of complexity $\mathcal{O}(|\mathcal{V}|^4)$, the optimal augmentation algorithm in [47] is of complexity $\mathcal{O}(|\mathcal{V}|^5)$, the simplified algorithm in [48] with complexity $\mathcal{O}(|\mathcal{V}||\mathcal{E}| + |\mathcal{V}|^2 \log|\mathcal{V}|)$, and the algorithm in [46] of complexity $\mathcal{O}(h|\mathcal{V}|^2)$with $h$ as the number of causalities, require higher complexity and running time than the proposed Algorithm 1.

[2] It is known that the problem of graph augmentation to reach $\kappa$-link-connectivity is generally NP-hard [42].

**Algorithm 1.** Design a κ-Link-Connected Network

**Data:** Seed graph $\mathcal{G}_W = (\mathcal{V}, \mathcal{E})$ and integer $\kappa$
**Result:** An augmented graph $\mathcal{G}'_W = (\mathcal{V}', \mathcal{E}')$ with at least $\kappa$-link-connectivity
Let $\mathcal{E}' \leftarrow \mathcal{E}$;
**For** *each node* $i \in \mathcal{V}$ **do**
Let $d_i \leftarrow \text{degree}(i)$ **if** $d_i < \kappa$ **then**
Find new $\kappa - d_i$ nodes $v_1, v_2, \dots, v_{\kappa - d_i}$ in $\mathcal{V}$;
**For** *each* $v_j$ **do**
**If** $(i, v_j) \notin \mathcal{E}'$ **then**
Add link $(i, v_j)$ to $\mathcal{E}'$;
Let $C \leftarrow \mathtt{ConnectedComponents}(\mathcal{G}_W)$;
**For** *each pair of components* $(C_i, C_j)$ *in* $C$ **do**
**If** *the components are not* $\kappa$*-connected* **then**
Add links between $C_i$ and $C_j$ to establish κ distinct links between the two components;
Ensure added links maintain $\kappa$-link-connectivity of the two components;
Update $\mathcal{E}'$ with the newly added links;
**Return** $\mathcal{G}'_W = (\mathcal{V}', \mathcal{E}')$

***Remark 2.*** *Node-connectivity metric of a graph is always less than the link-connectivity metric, therefore a* $\kappa$*-node-connected graph is always* κ*-link-connected, i.e., node-connectivity ≤ link-connectivity ≤ minimum degree following Whitney's theorem.*

If $\mathcal{G}_W$ is already given, the *min-cut algorithm* identifies the minimum number of links that, if removed, would disconnect the graph [49]. By identifying these links, additional links can be added to ensure that the minimum cut is greater than certain number. Moreover, *Menger's theorem* finds the relation between the connectivity of a graph and the maximum number of disjoint paths between two nodes [50]. Using this theorem, one can identify whether sufficient redundancy is present or needs to be introduced to meet the desired connectivity levels.

## 5. Illustrative Example and Simulations

The NCV and NCA models are commonly used in tracking and estimation problems, particularly in applications involving the prediction and filtering of unknown dynamic systems such as manoeuvring targets [33]. In the context of autonomous vehicles, self-driving cars and drones use these models to predict the motion of other vehicles to make safe navigation decisions, see [51-53]. For example, consider the AV to predict the future position and velocity of the HDVs to navigate through traffic safely. The vehicle uses the NCA model to estimate the state of the HDVs assuming that acceleration changes slowly and is subject to small random variations. This prediction helps in path planning, collision avoidance, and ensuring smooth and safe navigation in a dynamic traffic environment. In this direction, the HDV dynamics is modelled as (2) with the state $x = (\ddot{p_x}, \ddot{p_y}, \dot{p_x}, \dot{p_y}, p_x, p_y)^\top$ with $p_x, p_y$ as position of HDV in 2D, $\dot{p_x}, \dot{p_y}$ as its velocity, and $\ddot{p_x}, \ddot{p_y}$ as its acceleration. The input $u_k$ is considered random, and the system and input matrices are defined as,

$$A = \begin{bmatrix} 1 & 0 & 0 & 0 & 0 & 0 \\ 0 & 1 & 0 & 0 & 0 & 0 \\ T & 0 & 1 & 0 & 0 & 0 \\ 0 & T & 0 & 1 & 0 & 0 \\ \frac{T^2}{2} & 0 & T & 0 & 1 & 0 \\ 0 & \frac{T^2}{2} & 0 & T & 0 & 1 \end{bmatrix}, B = \begin{bmatrix} \frac{T^2}{2} & 0 \\ 0 & \frac{T^2}{2} \\ T & 0 \\ 0 & T \\ 1 & 0 \\ 0 & 1 \end{bmatrix}, \tag{8}$$

with parameter $T$ as the sampling time. We consider the network $\mathcal{G}_W$ given in Fig. 1 as the information-sharing network of $n = 4$ AVs. The entries of the associated adjacency matrix $W$ are set randomly such that row-stochasticity holds, i.e., $\sum_{j=1}^{n} W_{ij} = 1$. The shared information among the neighbouring AVs includes the state estimates and (possibly) measurements. Note that the AVs with no sensor measurement of the HDVs only share their estimates. For simulation, we set the sampling time as $T = 0.05$ and input $u$ as a uniform random process in the range $[-1,1]$. We adopt the distributed observer (5)-(6) and design the gain matrix $K_i$ with the Algorithm given in [37, Appendix]. The distributed observer enables every AV to locally estimate the state of the HDVs. Applying this distributed observer, the mean-square-error (MSE) at all AVs is shown in Fig. 3. Note that every vehicle $i$ only receives the information from the neighbouring vehicles $j \in N_i$, without receiving any direct message from the rest of the AVs. Despite this, it is clear from the figure that the MSE at all AVs is Schur (steady-state) stable under the proposed distributed observer design. This implies that every AV $i$ can track the state of every HDV, even those with no direct measurement by AV $i$. Recall that the steady-state error is due to noise considerations.

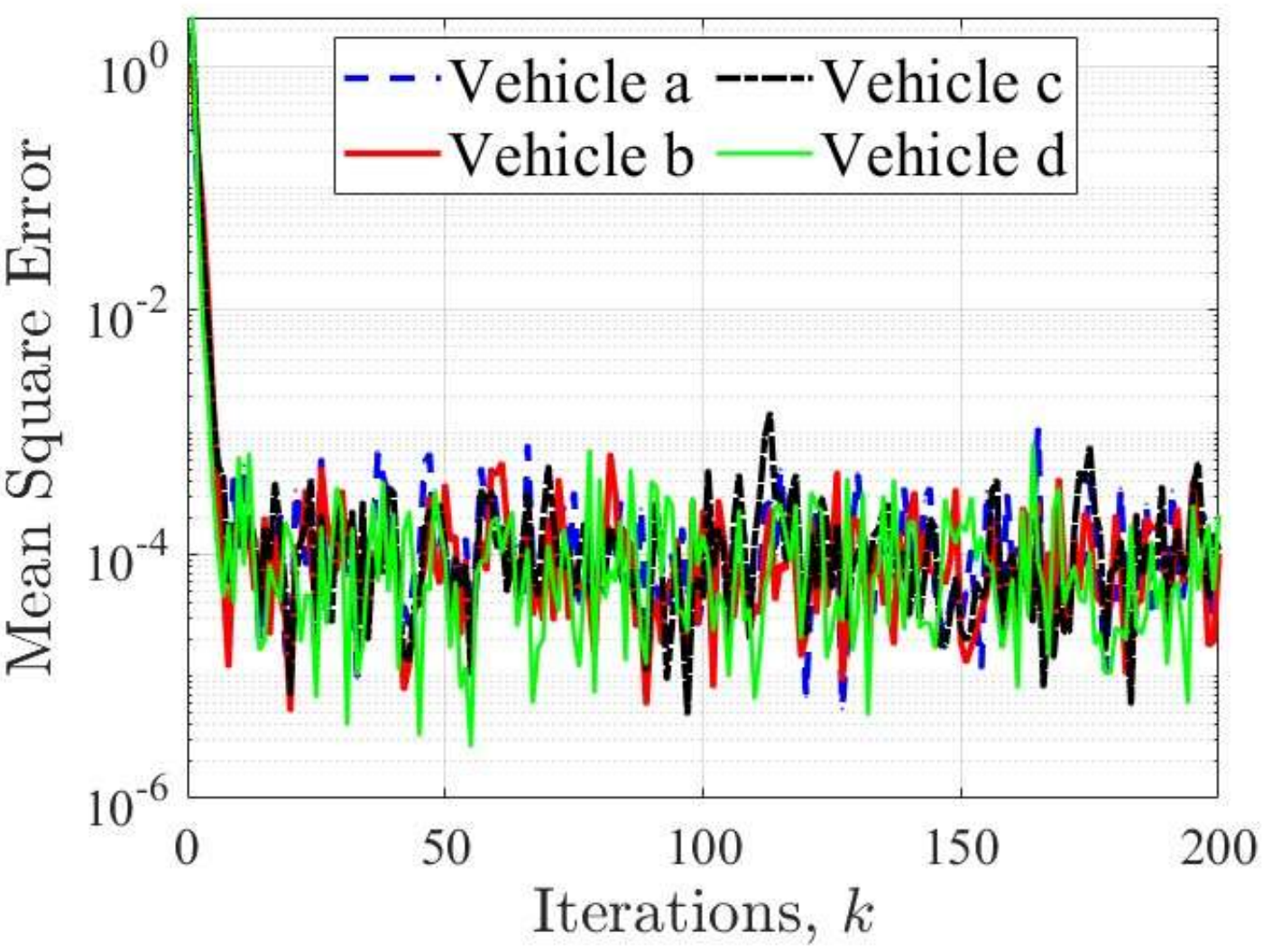

**Figure 3.** The time-evolution of the Mean-Square-Error (MSE) at all AVs connected via the network $\mathcal{G}_W$ shown in Fig. 1.

Next, we consider time-varying networks and evaluate the redundancy of the proposed observer design. Recall from Fig. 1 that the network $\mathcal{G}_W$ is resilient to removal (or failure) of AV '**b**' and the link '**bc**'. First, we remove the link '**bc**'. Note that, due to the existence of the redundant link '**ac**', the remaining network is still SC. Therefore, the condition in Theorem 1 holds, and the error dynamics is Schur stable. This is shown in Fig. 4, which represents stable MSE at all vehicles. Next, we consider the case that AV '**b**' is faulty and, therefore, removed from the network (its associated links are also removed). The remaining AVs and their reduced network $\mathcal{G}_W$ is still SC and satisfy the stability condition in Theorem 1; therefore, the MSE is stable at all remaining AVs, as illustrated in Fig. 5.

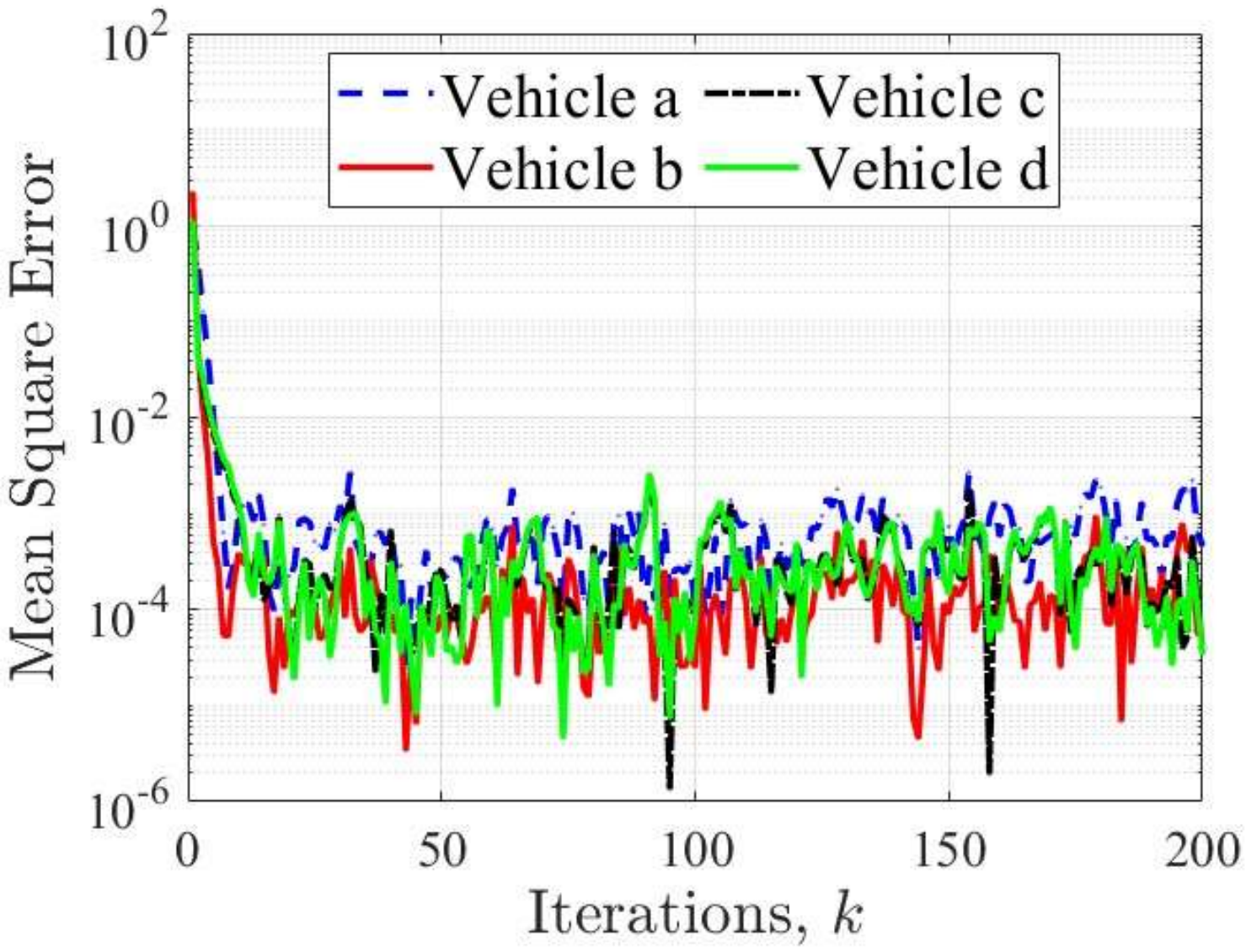

**Figure 4.** The time-evolution of the Mean-Square-Error (MSE) at all vehicles over the same network $\mathcal{G}_W$ in Fig. 1 after removing link '**bc**'.

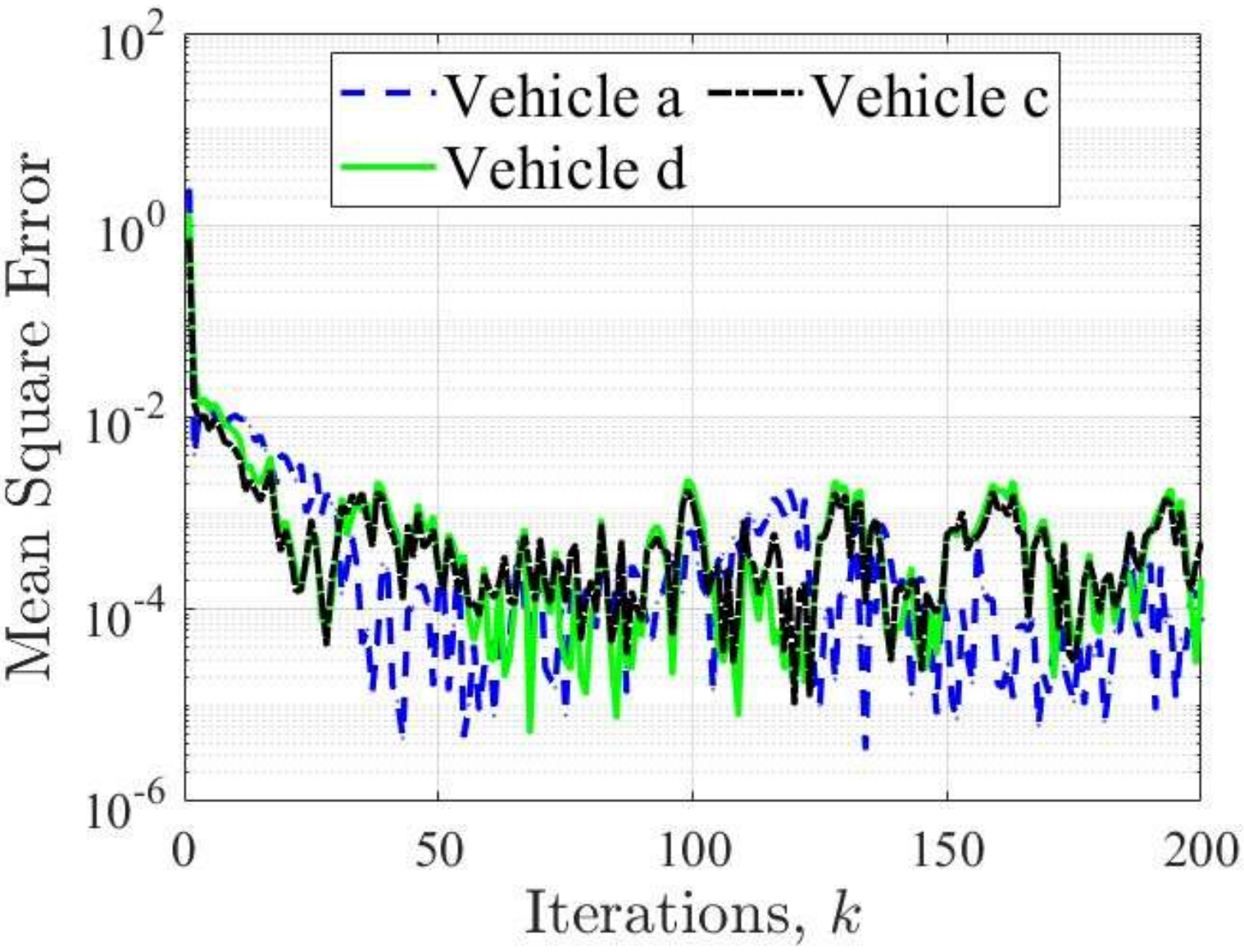

**Figure 5.** The time-evolution of the Mean-Square-Error (MSE) over the network $\mathcal{G}_W$ in Fig. 1 after removing AV '**b**'.

***Remark 3.*** *Recall that, although the algorithm is illustrated over a small-scale vehicular network, it can be efficiently extended to larger scale setups. This is because the proposed algorithm is of* *complexity order $\mathcal{O}(|\mathcal{V}|^2)$; this implies that the complexity of the proposed algorithm can be extended polynomially to larger scale with efficient computations.*

## 6. Conclusion

In this paper, providing the graph-theoretic formulation for observability, the merging scenario for the network of AVs and HDVs is presented aiming to make the dynamics of all HDVs observable and trackable by all the AVs. This allows the AVs to estimate the state of all HDVs by applying relevant distributed estimation protocols for safe mixing traffic strategies. Then, an efficient strategy is proposed to add redundancy in the network of vehicles for resilient design against failures. This helps to estimate the (position, velocity, acceleration) state of HDVs even in the case of link failure due to unreliable communications or when the vehicles go out of communication range of each other. This paper may have many applications in ITS including in autonomous vehicle coordination. As future research direction, one can incorporate real-world constraints in the data-sharing network of vehicles, such as quantization to address limitations in network traffic.

Future research may connect the proposed framework with empirical research fields where multiple sensing technologies are already integrated to analyze real driving behavior. For example, naturalistic driving studies combine GNSS positioning, onboard sensors, and vehicle telemetry to capture real-world driving dynamics [54-56]. Within this context, the concept of GNSS signal integrity has been highlighted as a key challenge, since positioning signals alone may not always provide reliable information and therefore often need to be complemented with other sensing technologies. This line of research could help place the proposed theoretical framework within a broader context of multi-sensor transportation systems and data integration challenges.

**Author Contributions:** Conceptualization, M.D.; methodology, M.D.; software, M.D.; validation, M.D.; formal analysis, M.D.; investigation, M.D.; resources, M.D.; writing—original draft preparation, M.D.; writing—review and editing, M.D.; visualization, M.D; supervision, M.D.; project administration, M.D.

**Funding:** This work is funded by Semnan University, research grant No. 226/1403/1403208.

**Data Availability Statement:** No new data were created or analyzed in this study. The MATLAB simulation code and relevant configuration files used to generate the results are available from the authors upon request.

**Conflicts of Interest:** The authors declare no conflicts of interest.